\documentclass[
aps,%
11pt,%
final,%
notitlepage,%
oneside,%
twocolumn,%
nobibnotes,%
nofootinbib,%
superscriptaddress,%
noshowpacs,%
centertags]%
{revtex4}
\renewcommand{\baselinestretch}{0.9}
\begin{document}

\title{New Low Surface Brightness Dwarf Galaxies Detected~Around~Nearby~Spirals}

\author{\firstname{I.~D.}~\surname{Karachentsev}}
\email{ikar@sao.ru}\affiliation{\saoname}

\author{\firstname{P.}~\surname{Riepe}}
\author{\firstname{T.}~\surname{Zilch}}
\author{\firstname{M.}~\surname{Blauensteiner}}
\author{\firstname{M.}~\surname{Elvov}}
\author{\firstname{P.}~\surname{Hochleitner}}
\author{\firstname{B.}~\surname{Hubl}}
\author{\firstname{G.}~\surname{Kerschhuber}}
\author{\firstname{S.}~\surname{K\"{u}ppers}}
\author{\firstname{F.}~\surname{Neyer}}
\author{\firstname{R.}~\surname{P\"{o}lzl}}
\author{\firstname{P.}~\surname{Remmel}}
\author{\firstname{O.}~\surname{Schneider}}
\author{\firstname{R.}~\surname{Sparenberg}}
\author{\firstname{U.}~\surname{Trulson}}
\author{\firstname{G.}~\surname{Willems}}
\author{\firstname{H.}~\surname{Ziegler}}
\affiliation{Tief Belichtete Galaxien group of Vereinigung der
Sternfreunde e.V., Fachgruppe Astrofotografie,\\ Heppenheim,
D-64629 Germany}

\received{June 22, 2015}  \revised{July 31, 2015}

\begin{abstract}
We conduct a survey of low surface brightness (LSB) satellite
galaxies around the Local Volume massive spirals using long
exposures with small amateur telescopes. We identified 27 low and
very low surface brightness objects around the galaxies NGC\,672,
891, 1156, 2683, 3344, 4258, 4618, 4631, and~5457 situated within
10~Mpc from us, and found nothing new around NGC\,2903, 3239,
4214, and 5585. Assuming that the dwarf candidates are the
satellites of the neighboring luminous galaxies, their absolute
magnitudes are in the range of \mbox{$-8.6 > M_B > -13.3$}, their
effective diameters are 0.4--4.7~kpc, and the average surface
brightness is $26\fm1/\sq\arcsec$. The mean linear projected
separation of the satellite candidates from the host galaxies is
73~kpc. Our spectroscopic observations of two LSB dwarfs with the
Russian 6-meter telescope confirm their physical connection to the
host galaxies NGC\,891 and NGC\,2683.
\end{abstract}

\maketitle

\section{INTRODUCTION}

The two past decades witnessed the rapid formation of a new
direction in extragalactic astronomy, called the cosmology of the
nearby universe. This is facilitated by the ongoing wide-field
surveys of the northern and southern  sky in the optical,
infrared, and radio
ranges~\mbox{\cite{abaz2009:Karachentsev_n,gio2005:Karachentsev_n,hay2011:Karachentsev_n,jar2000:Karachentsev_n,jar2003:Karachentsev_n,kor2004:Karachentsev_n,zwa2003:Karachentsev_n,kov2009:Karachentsev_n,ton2012:Karachentsev_n}},
which allowed astronomers to detect a lot of dwarf galaxies and to
measure their radial
velocities~\mbox{\cite{kara1998:Karachentsev_n,kara2000:Karachentsev_n,huc2009:Karachentsev_n}}.
Accurate bulk measurements of distances to nearby galaxies carried
out with the
 Hubble Space Telescope  are also a significant advancement factor.
According to the latest  observational data report,  titled the
{\it Updated Nearby Galaxy Catalog}
(UNGC)~\cite{kar2013:Karachentsev_n}, the surrounding  volume
around   the Milky Way with a  radius of about 10~Mpc contains
about nine hundred galaxies, the majority of which have their
distances and radial velocities measured, and stellar masses, star
formation rates, and other basic characteristics determined. The
generally accessible database of observational data on the
galaxies of the Local Volume~\cite{kai2012:Karachentsev_n} ({\tt
http://www.sao.ru/lv/lvgdb/}) is  regularly updated with new
objects. About 85\% of the UNGC~\cite{kar2013:Karachentsev_n}
sample are dwarf galaxies whose integrated  luminosity is lower
than that of the Magellanic Clouds.

The numerical simulations of the large-scale structure of the
Universe performed on supercomputers within the standard
$\Lambda$CDM cosmological
model~\cite{moo1999:Karachentsev_n,kly1999:Karachentsev_n}
revealed a huge discrepancy in the observed number of dwarf
galaxies relative to their number expected  within the standard
model. The observed number of dwarf satellites around nearby high
luminosity galaxies turned out to be ten times smaller than
expected. This as yet unexplained situation was called the
``missing satellites'' paradox. An ad hoc search for dwarf
galaxies around the nearest massive galaxies
M\,31~\cite{iba2007:Karachentsev_n,iba2013a:Karachentsev_n,mar2009:Karachentsev_n}
and M\,81~\cite{chi2009:Karachentsev_n,chi2013:Karachentsev_n} has
only partially alleviated this paradox. Therefore, a further
in-depth search for yet fainter dwarf systems
 continues to be a topical task of observational cosmology of the nearby universe.

According to theoretical and observational data, the integrated
luminosity of a galaxy $L$ is proportional to the cube of its
effective linear diameter $A$. This implies that the average
luminosity density of galaxies $L/A^3$ is about the same
regardless of their size, however, their average surface
brightness   $L/A^2$ drops with decreasing linear diameter of the
galaxy. For this reason the smallest dwarf satellites are to be
searched for among LSB objects. Medium-sized amateur telescopes
(about 0.3~m in diameter) with a focal ratio
\mbox{$f/D\sim4$--$8$} and equipped with  CCD detectors can be
used for this task. The images obtained on such telescopes with
exposures of about 10~hours reveal quite well objects with the
surface brightness \mbox{${\rm SB}\sim27^{\rm m}$--$28^{\rm
m}/\sq\arcsec$} and angular sizes exceeding $0\farcm2$, which
roughly corresponds to the typical parameters of dwarf galaxies
located within the Local Volume.

According to preliminary estimates, the UNGC  is about 50\%
complete for the galaxies with the absolute magnitude $M_B$
brighter than $-11\fm0$ and the surface brightness ${\rm
SB}\leq26^{\rm m}/\sq\arcsec$. The characteristic linear diameter
of dwarf galaxies near the 50\% limit is about 1~kpc, which
corresponds to an angular size
 $a\sim0\farcm3$  at the far edge of the Local Volume.
Therefore, hours-long exposures of the vicinity of nearby bright
galaxies with modern amateur telescopes allow successful detection
of new dwarf satellites around them. Such a systematic
observational program provides an independent opportunity to
clarify the extent of completeness of the UNGC in terms of the
luminosity of dwarf galaxies, their linear dimensions and surface
brightnesses.

\section{TBG GROUP OBSERVATIONAL PROJECT}

The TBG  (Tief Belichtete Galaxien) group deals with very long
exposure images of galaxies  made using amateur telescopes of
medium caliber. The group was organized by P.~Riepe in January
2012 in the {\it Astrophotography} department of the German
association VdS  (Vereinigung der Sternfreunde e.V.). Now the
group consists of about 30 astrophotographers from Germany,
 Austria, and Switzerland, equipped with 10~to 110~cm diameter telescopes. Some TBG telescopes are located in the United States
and Spain and are  remotely operated. The entire project is
coordinated by P.~Riepe and T.~Zilch. For deep-sky images,
reaching \mbox{${\rm SB}\sim28^{\rm m}/\sq\arcsec$}, the group
uses high-quality CCD detectors and data reduction packages which
include the dark frame subtraction, flat field correction, and
calibration procedures.

One of the main tasks of the TBG group  is photographing the
neighborhoods of nearby bright galaxies to look for LSB dwarf
satellites around them. The program includes more than 50~galaxies
of sufficiently high luminosity located within a distance of
10~Mpc. Exposure time with broadband filters varied depending on
weather conditions. The typical exposure lasted from 10 to
15~hours, though in some cases it reached up to 50~hours.


After the necessary image processing steps  and contrasting of the
images, we conducted a visual search for low and very low surface
brightness objects, focusing on the average characteristics of the
known satellites of the Milky Way and Andromeda (M\,31).


Below we present the results of the satellite candidate search
around   thirteen  nearby spiral galaxies, obtained in
implementing the initial phase of this program. The following
articles of this series will present the results of monitoring the
neighborhoods of still about 40 massive galaxies in the Local
Volume.

     \section{FIRST-SEASON OBSERVATIONS}

     \subsection{NGC\,4631}

This  late-type spiral galaxy, seen edge-on, is the brightest
representative of a scattered group, which includes about
30~members. The distance to it, $D=7.38$~Mpc, was measured by
Radburn-Smith et~al.~\cite{rad2011:Karachentsev_n} from the
luminosity of red giant branch stars. Another bright spiral
galaxy, NGC\,4656, is located at an angular distance of
$32\arcmin$ south of NGC\,4631. Both of them have  structural
distortions which are obviously caused by  mutual gravitational
perturbations.

In  spring 2013, several  TBG group members have obtained the
images of NGC\,4631 and its environs  with a total exposure of
$24^{\rm h}$. Methodological details of these observations were
given earlier~\cite{kar2014a:Karachentsev_n}.  We marked three
very low surface brightness dwarf galaxies  in the combined image,
dw1, dw2, dw3, as well as a tidal strip that stretches from
NGC\,4631 to the northwest through dw1 and to the southeast
towards the galaxy NGC\,4656. The presence of this tidal strip was
later confirmed by Martinez-Delgado
et~al.~\cite{mar2014:Karachentsev_n}. In February 2013 and March
2014, F.~Neyer obtained new, deeper images of the NGC\,4531/56
pair with an  exposure of $49\fh5$
  (Fig.~1,  see also {\tt http://tbg.vdsastro.de}). This
image reveals different signs of interaction between  the
components of this pair  with a characteristic surface brightness
of $29^{\rm m}$--$30^{\rm m}/\sq\arcsec$. Some of them are
indistinguishable from faint reflection nebulae. All the three
objects dw1/2/3, according to their texture and location, are
probably the physical satellites of NGC\,4631.   However, radial
velocity measurements of these dwarfs are required to confirm this
obvious assumption,  which is a difficult observational task.


      \subsection{M\,101 $=$ NGC\,5457}

The spiral galaxy M\,101, viewed face-on, is one of the most
prominent representatives of the Local Volume population. The
distance to it by the Cepheids is estimated as
7.38~Mpc~\cite{fer2000a:Karachentsev_n},
  which happens to coincide with the distance to NGC\,4631. According to~\cite{kar2014b:Karachentsev_n},
  M\,101 has six satellites, including an intergalactic  H\,I~cloud GBT\,1355+5439~\cite{mih2012:Karachentsev_n}.
In 2009 and 2010 Mihos et~al.~\cite{mih2013:Karachentsev_n}
obtained deep images of the  neighborhood of M\,101 with the
Burrell Schmidt telescope. The image mosaics covered a field sized
\mbox{$2\fdg5\times 2\fdg5$}  with  the surface brightness limit
 \mbox{${\rm SB}(B)\sim29\fm5/\sq\arcsec$}. The authors noted the
presence in M\,101 of two structural disturbances on the
periphery: the northeastern  spot and the eastern protrusion.
However, they did not report a discovery of any new nearby
satellites.  In March 2012, a deep image of M\,101 with an
exposure of $40^{\rm h}$ was obtained by F.~Neyer using a
telescope with a diameter of 15~cm and a focal ratio of~$f/7.2$.
The image size is $121\arcmin\times 80\arcmin$. The reproduction
of the fragments of this image  is shown in
 three panels of Fig.~2. In addition to the above-mentioned distortions of the periphery of
M\,101, on the
  northern and northeastern side  the image reveals   ten small LSB objects.
All of them are located in the eastern half of the image with
respect to M\,101, which looks pretty mysterious.

In May--June 2013, van~Dokkum et~al.~\cite{dok2014:Karachentsev_n}
and Merritt et~al.~\cite{mer2014:Karachentsev_n} performed a
survey of the neighborhood of M\,101 using  the Dragonfly
Telephoto Array. This telescope is a robotic system of eight
lenses, each one with a focal length of 40~cm and a focal ratio of
$f/2.8$. This results in an effective aperture of 403~mm and the
summed up focal ratio of about $f/1$. The field of view is
\mbox{$2\fdg6\times 1\fdg9$} with a resolution  of $2\farcs8$/px.
Given a total exposure of $ 35^{\rm h}$, the Dragonfly team
reached a surface brightness limit \mbox{${\rm
SB}\sim29\fm5/\sq\arcsec$}.
 As a result, the authors found seven M\,101 dwarf
satellite candidates and designated them as  DF1, DF2,~...,~DF7.
Six of them proved to be common with the objects in the image of
F.~Neyer, and one, DF5, is located outside the image field. At the
same time, two LSB objects that we have designated as M\,101\,dwA
and M\,101\,dwC were not noted by the Dragonfly team. The
Dragonfly deep   image limit  allowed Merritt
et~al.~\cite{mer2014:Karachentsev_n} to conduct surface photometry
of the detected objects and to determine their integrated $g$ and
$r$ magnitudes, effective diameter,  central surface brightness,
and the S\'ersic structural index. Recently, M.~Elvov obtained a
broader image of the vicinity of M\,101, where we found two more
satellite candidates: M\,101\,dwB and M\,101\,dwD.

\begin{figure}
 \setcaptionmargin{5mm} \onelinecaptionsfalse \captionstyle{normal}
 \vspace{2mm}
\includegraphics[width=\columnwidth]{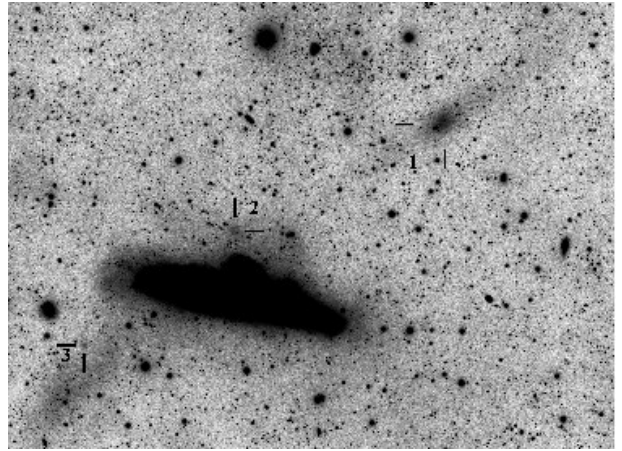}
\caption{The  galaxy NGC\,4631 and three of its supposed
companions, located along the diagonal tidal  stripe. A
$45\arcmin\times 33\arcmin$ fragment of an image obtained by
F.~Neyer with an exposure time of $49\fh5$. North is at the top,
east is to the left.}
\end{figure}

\begin{figure*}
 \setcaptionmargin{5mm} \onelinecaptionsfalse \captionstyle{normal}
 \vspace{2mm}
 \includegraphics[width=0.48\textwidth]{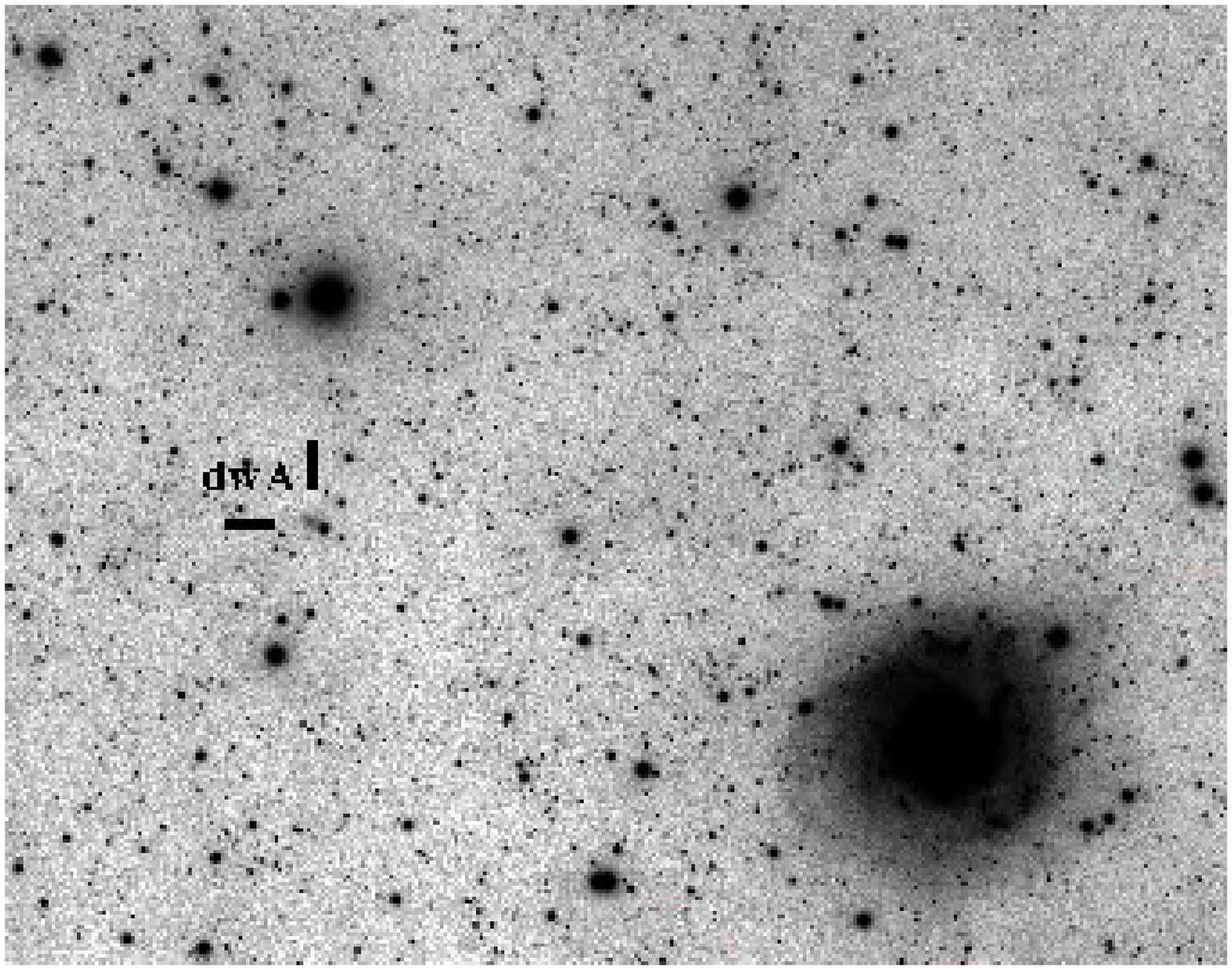}\\
 (a)\\
 \vspace{2mm}
\includegraphics[width=0.52\textwidth]{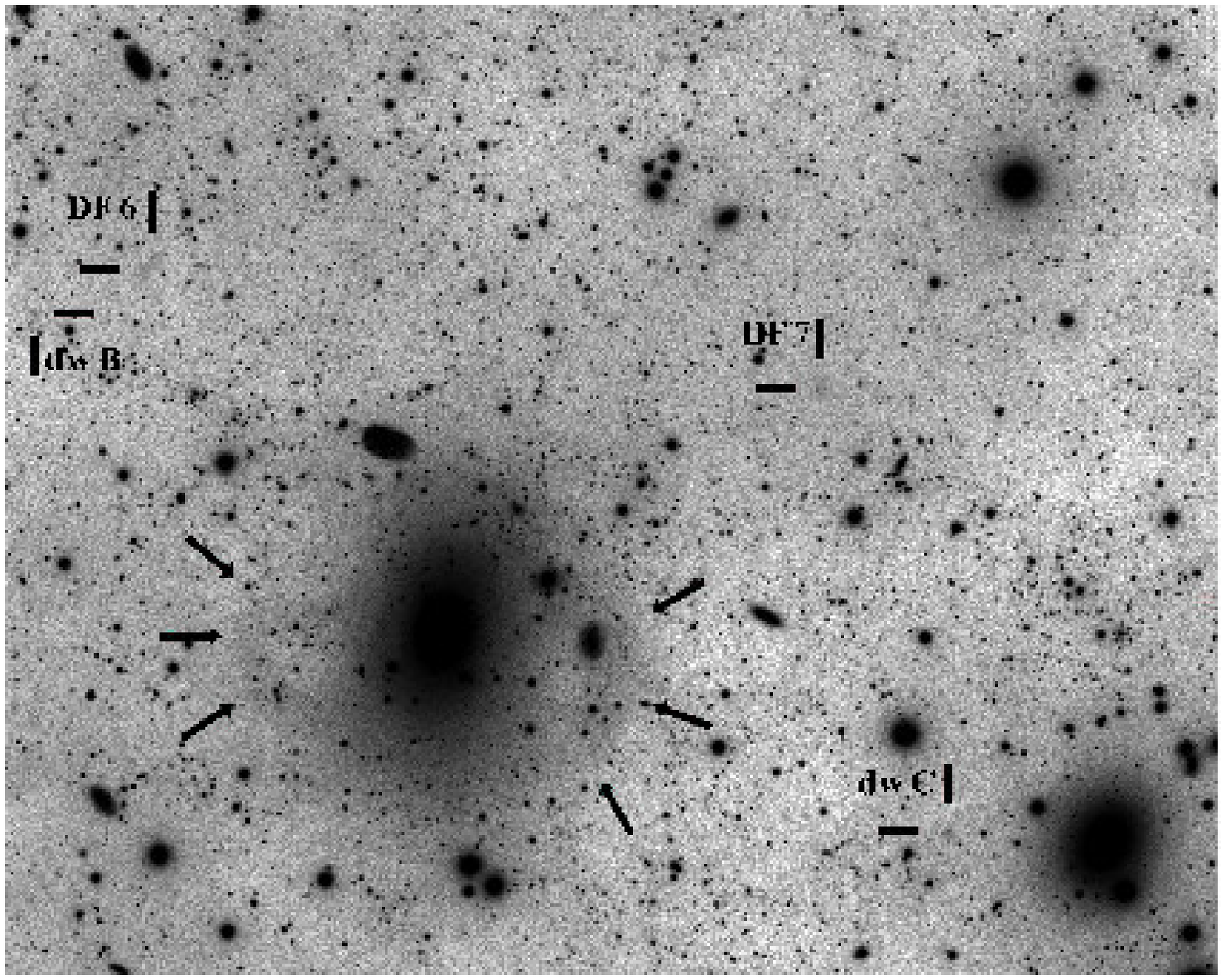}\hfill\includegraphics[width=0.46\textwidth]{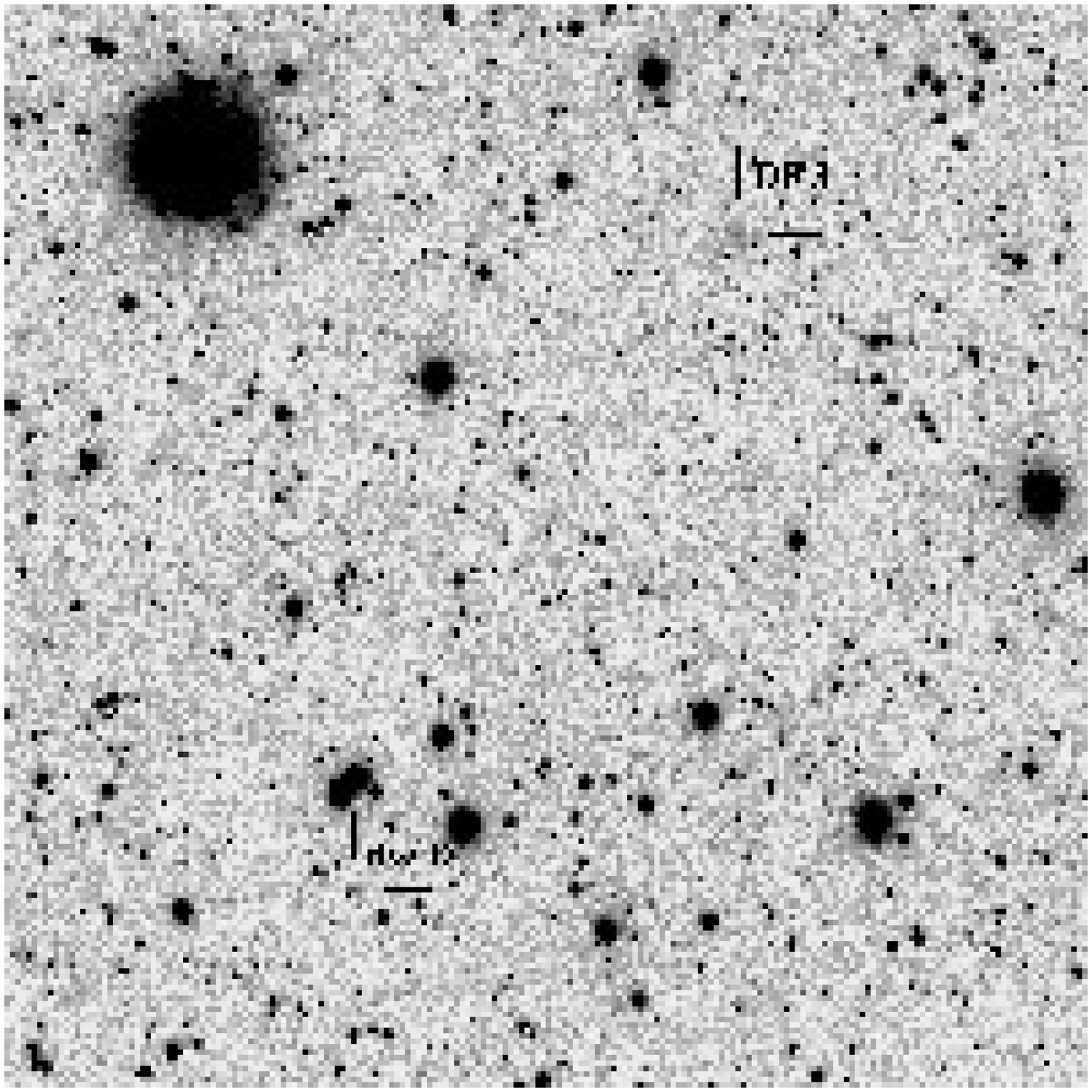}\\
 \hspace{5mm}(b)\hspace{85mm}(c)
 \caption{Fragments of the image displaying the neighborhood of M\,101, obtained by
F.~Neyer with an exposure of $45\fh6$. (a)~The region of an M\,101
satellite, the galaxy NGC\,5477, and a new satellite candidate
dwA; the frame size is $36\arcmin\times 24\arcmin$. (b) The area
to the northeast of M\,101 sized $39\arcmin\times 31\arcmin$, in
which the   Dragonfly~DF\,6 and DF\,7 objects  and two new M\,101
 satellite candidates dwB and dwC are located.
Down from the center, a far background galaxy  NGC\,5485 is
located (a distance of~28~Mpc), around which  diffuse elliptical
shells, shown by the arrows were detected for the first time. (c)
A fragment of an image obtained by M.~Elvov with an exposure of
$11\fh3$. In the area sized $34\arcmin\times 34\arcmin$, the
 NGC\,5477 galaxy  (the  upper left corner), the
Dragonfly~DF\,3 object, and a new  M\,101 satellite candidate dwD
are visible. In all the images north is at the top, and east is to
the left.}
\end{figure*}

\begin{figure}
 \setcaptionmargin{5mm} \onelinecaptionsfalse \captionstyle{normal}
 \vspace{2mm}
\includegraphics[width=\columnwidth]{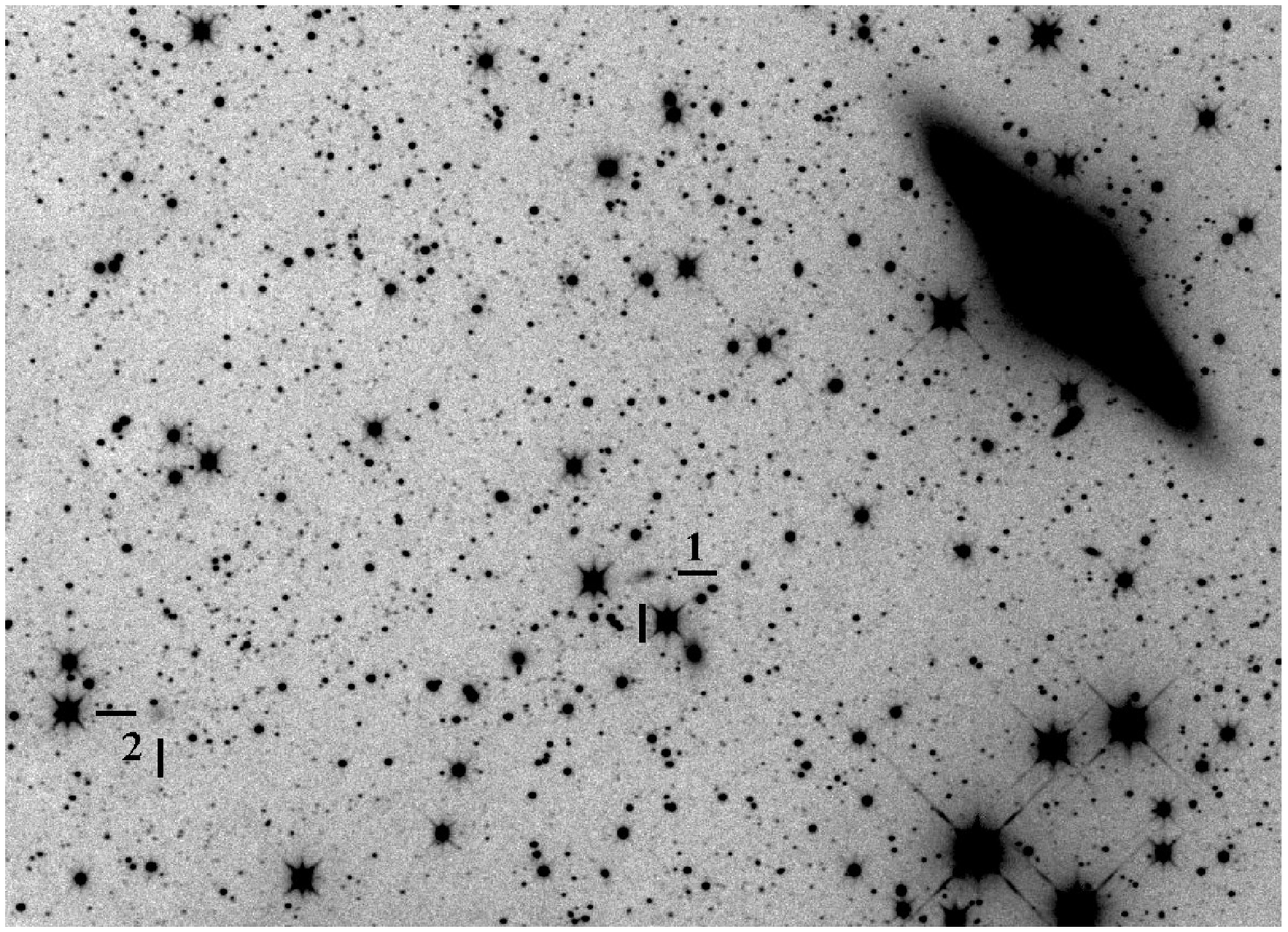}
\caption{An image of the spiral galaxy NGC\,2683 (in the upper
right corner) and its two prospective satellites. The frame size
is $29\arcmin\times 21\arcmin$. The image was obtained by
G.~Kerschhuber with an exposure of $12^{\rm h}$. Based on the
measured radial velocity, the object~``1'' is a true satellite of
NGC\,2683.}
\end{figure}

     \subsection{NGC\,2683}

This   massive spiral galaxy at a distance of\linebreak
\mbox{9.36~Mpc~\cite{kar2015:Karachentsev_n}}
  has two close and one distant satellite, KK\,69, KK\,70, and AGC\,182595 respectively, the distances to which were measured with the
Hubble Space
Telescope~\cite{mcq2014:Karachentsev_n,kar2015:Karachentsev_n}.
This poor group is located in the region of low galaxy number
density
 on the front edge of the  \mbox{Gemini--Leo} Void.


Two deep images centered on NGC\,2683 were obtained by M.~Elvov
and  G.~Kersch\-huber.  The field sizes  and exposures were
\mbox{$31\arcmin\times 22\arcmin$,} $t=15^{\rm h}$ and
\mbox{$41\arcmin\times 55\arcmin$,}  $t=12^{\rm h}$ respectively.
The reproduction of the second image fragment (Fig.~3) reveals two
LSB dwarf objects, designated as dw1 and dw2. The brighter of them
is identified with a UV source of the GALEX
survey~\cite{gil2007:Karachentsev_n}. Its image in the H$\alpha$
line~\cite{kar2015a:Karachentsev_n}, taken with the 6-meter
telescope of the Special Astrophysical Observatory of the Russian
Academy of Sciences (SAO RAS) with the \mbox{SCORPIO} focal
 reducer~\cite{afa2005:Karachentsev_n}, detects weak emission.
The [O\,III], H$\beta$, and H$\gamma$~\cite{kar15:Karachentsev_n}
 emission lines are
present in the spectrum of the object obtained with the same
telescope~\cite{kar15:Karachentsev_n}.  Based on them, the
heliocentric radial velocity of the object  is
\mbox{$V_h=380\pm25$~km\,s$^{-1}$}, which is close to the radial
velocity of  NGC\,2683 itself, \mbox{$V_h=411\pm4$~km\,s$^{-1}$}.
The agreement of radial velocities indicates that the new dwarf
galaxy is a real companion of NGC\,2683. Low surface brightness of
the second (dSph) dwarf galaxy makes
  its radial velocity measurement quite difficult.

\begin{figure}
 \setcaptionmargin{5mm} \onelinecaptionsfalse \captionstyle{normal}
 \vspace{2mm}
\includegraphics[width=\columnwidth]{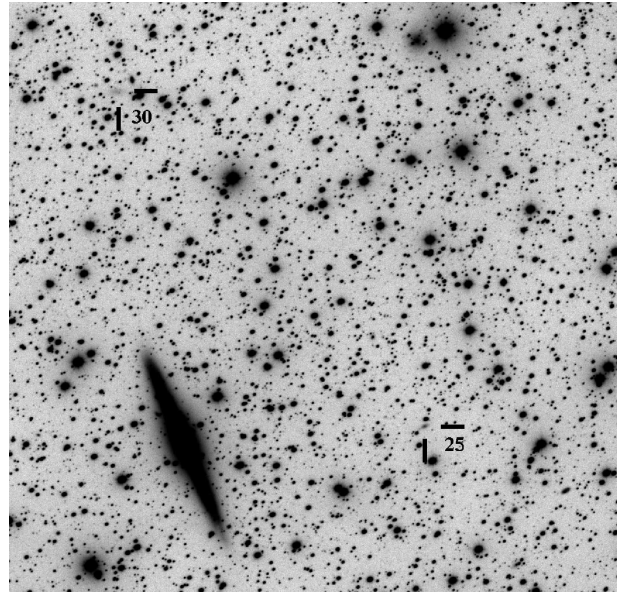}\\
 (a)\\
 \vspace{2mm}
\includegraphics[scale=0.4,bb=105 35 400 325,clip]{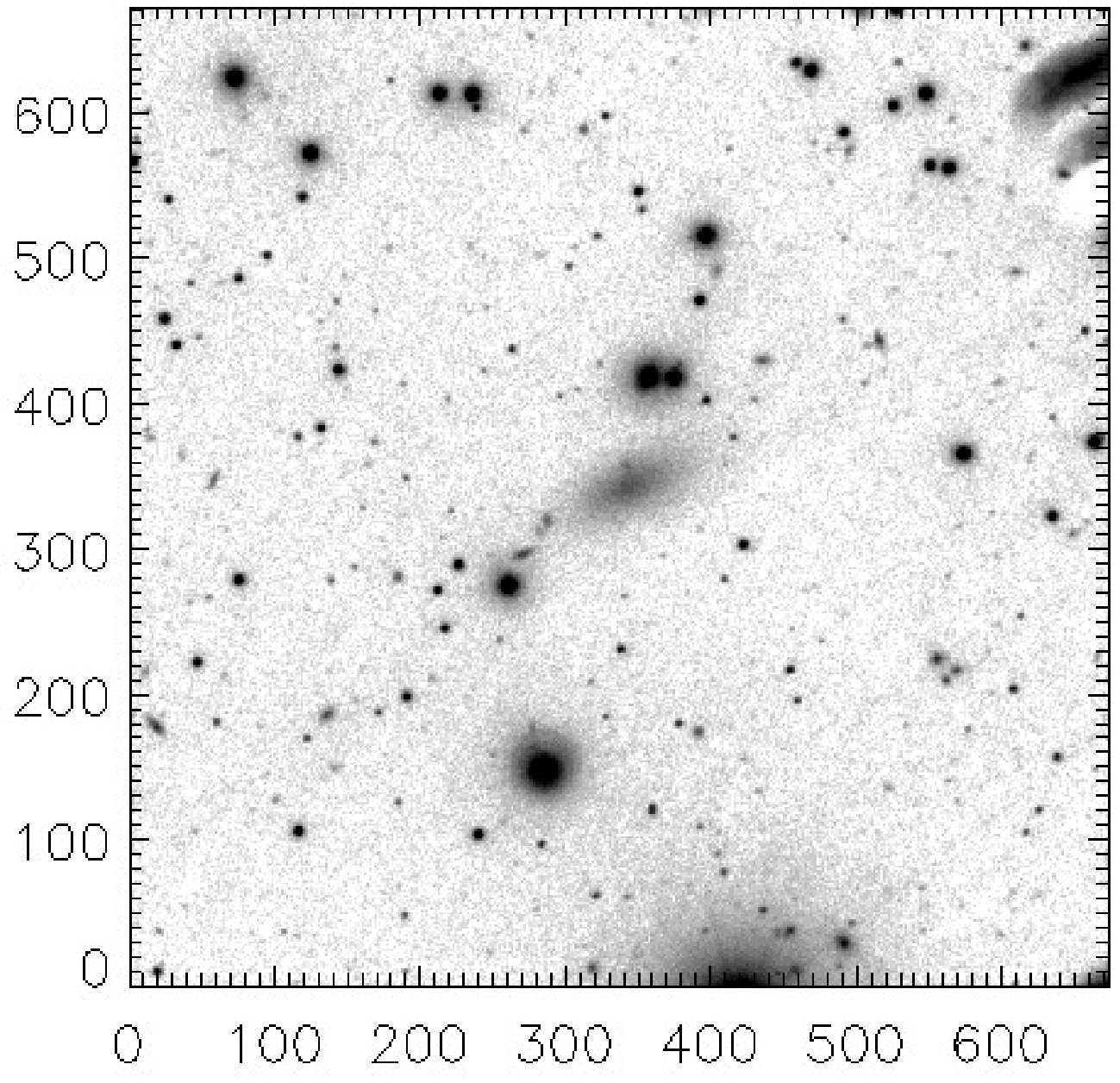}
\includegraphics[scale=0.4,bb=105 35 400 325,clip]{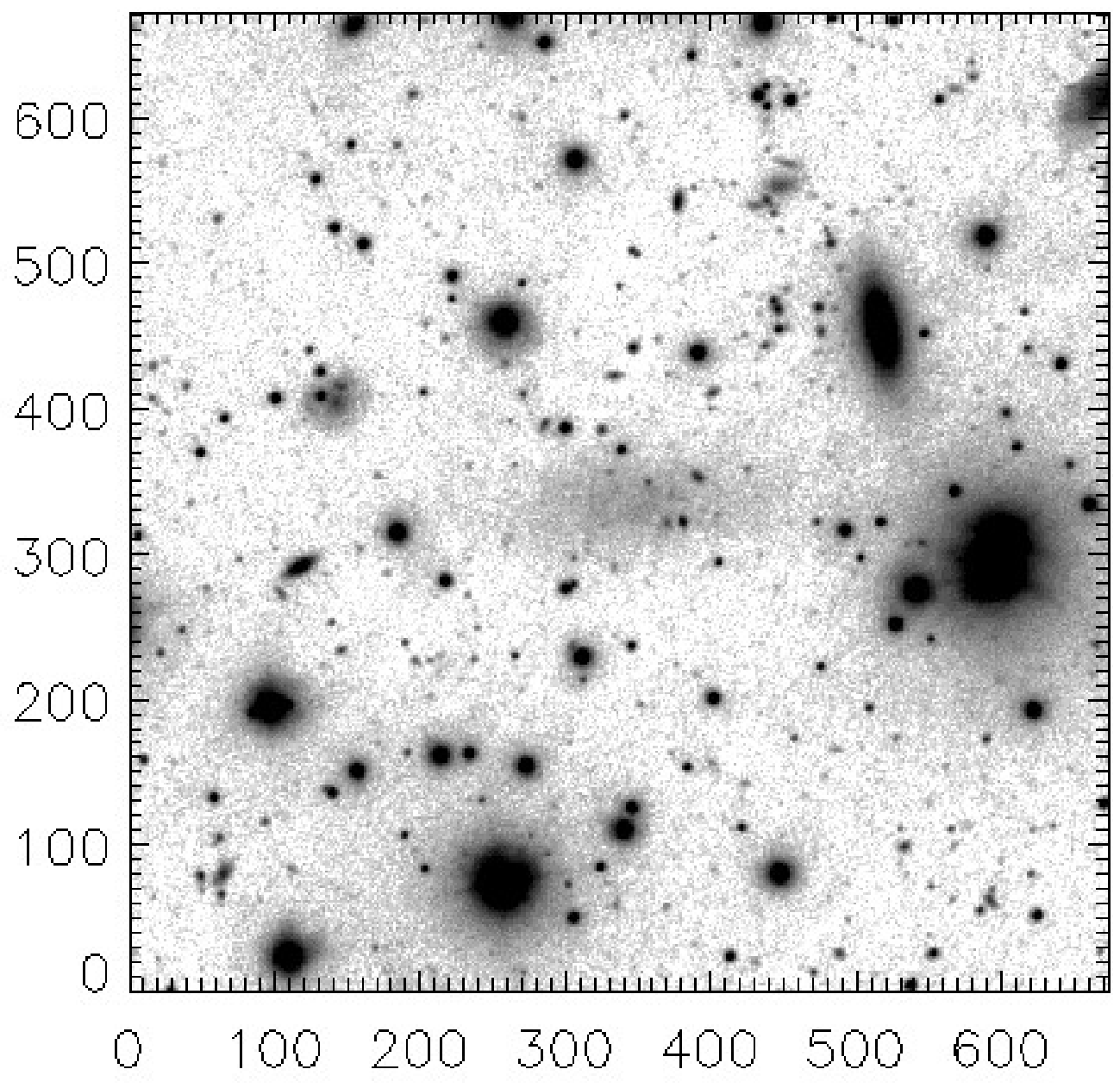}\\
(b)\hspace{40mm}(c)
 \caption{(a) The spiral galaxy NGC\,891 and its two
dwarf satellites: dwA~$=$ [TT\,2009]\,25 and dwB~$=$
[TT\,2009]\,30. A $33\arcmin\times 32\arcmin$ fragment of an image
obtained by M.~Elvov with an exposure of $12^{\rm h}$.
(b,~c)~Images of the dwarf galaxies [TT\,2009]\,25 and
[TT\,2009]\,30, obtained by S.~S.~Kaisin on the SAO RAS 6-meter
 telescope with the SED607 filter with 1650-s and 2400-s  exposures respectively.
In all the images north is at the top, east is to the left.}
\end{figure}

     \subsection{NGC\,891}

According to~\cite{kai2012:Karachentsev_n}, the entourage of the
spiral galaxy NGC\,~891, seen edge-on, consists of  four Sm and
Irr galaxies: DDO\,22, DDO\,24, UGC\,1807, and UGC\,2172. The
distance to the main galaxy,  9.77~Mpc, was determined from the
surface brightness fluctuations~\cite{fer2000b:Karachentsev_n}.
The group itself is associated with the spiral galaxies NGC\,925
and NGC\,1023, forming a more extended and
  non-virialized complex.

\begin{figure}
 \setcaptionmargin{5mm} \onelinecaptionsfalse \captionstyle{normal}
 \vspace{2mm}
\includegraphics[width=\columnwidth]{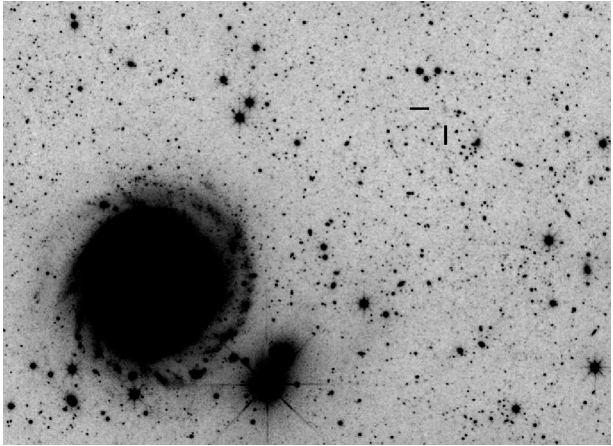}
\caption{An image of the spiral galaxy NGC\,3344 and its suspected
dwarf companion NGC\,3444\,dw1, marked by arrows. The image was
obtained by G.~Kerschhuber and  M.~Blauensteiner based on
simultaneous data from  four telescopes with a total exposure of
$28\fh8$. The size of the fragment is $22\arcmin\times 16\arcmin$.
North is at the top, east is to the left.}
\end{figure}

Faint satellites of NGC\,891 were searched for by Trentham and
Tully~\cite{tre2009:Karachentsev_n} and
Schulz~\cite{sch2014:Karachentsev_n}. The former authors used
broadband exposures with the MegaCam detector of the 3.6-m CFHT
telescope. In the immediate surroundings of NGC\,891, the authors
found only two dwarf companion candidates: [TT\,2009]\,25 and
[TT\,2009]\,30. Schulz~\cite{sch2014:Karachentsev_n}
 searched for new satellites of NGC\,891 based on the  data from
various available surveys of the sky from the ultraviolet (GALEX)
to the infrared (2MASS, WISE) range. Two out of  seven suspected
new satellites  turned out to be  already known, and the remaining
five are likely to be the objects of  distant background. A CCD
image of the vicinity of NGC\,891 obtained by M.~Elvov on the
\mbox{10-cm} refractor $(f/5$)
 with an exposure of $12^{\rm h}$ (Fig.~4a) detects the presence of
  two LSB dwarf galaxies dwA and dwB, which are identified with the objects [TT\,2009]\,25 and [TT\,2009]\,30.
We have obtained large-scale images of both galaxies in the
H$\alpha$ line and in the continuum (filter
SED607)~\cite{kar2015:Karachentsev_n} with the SAO RAS  6-m
telescope (Figs.~4b and~4c).  They did not reveal any H$\alpha$
emission,
 even though in the GALEX  survey they show weak FUV fluxes.
 For a brighter spheroidal galaxy [TT\,2009]\,25,
a spectrum was obtained with the SAO RAS 6-m
telescope~\cite{kar15:Karachentsev_n}. Based on three absorption
lines, the radial velocity of the galaxy is
 $V_h=692\pm58$~km\,s$^{-1}$,  which is close to the heliocentric velocity
 of   NGC\,861 itself: \mbox{$V_h=526\pm7$~km\,s~$^{-1}$.}
According to the texture of the object [TT\,2009]\,30, it too,
with a high probability, is
  a satellite of NGC\,891.


       \subsection{NGC\,3344 $=$ KIG\,435}

This is an isolated galaxy seen face-on, with a regular spiral
pattern. None of its physical satellites were known to date. An
image of the surroundings of NGC\,3344  made by Kerschhuber and
Blauensteiner with a total exposure of $28\fh8$ and field of
\mbox{$40\arcmin\times 30\arcmin$} reveals a suspected LSB
satellite
 (Fig.~5) $12\arcmin$ northwest of NGC\,3344. In this place, the
SDSS survey~\cite{abaz2009:Karachentsev_n} reveals a faint bluish
spot with an angular size of $0\farcm3$. The GALEX  survey did not
detect this object.

\begin{figure*}
 \setcaptionmargin{5mm} \onelinecaptionsfalse \captionstyle{normal}
 \vspace{2mm}
\includegraphics[width=\textwidth]{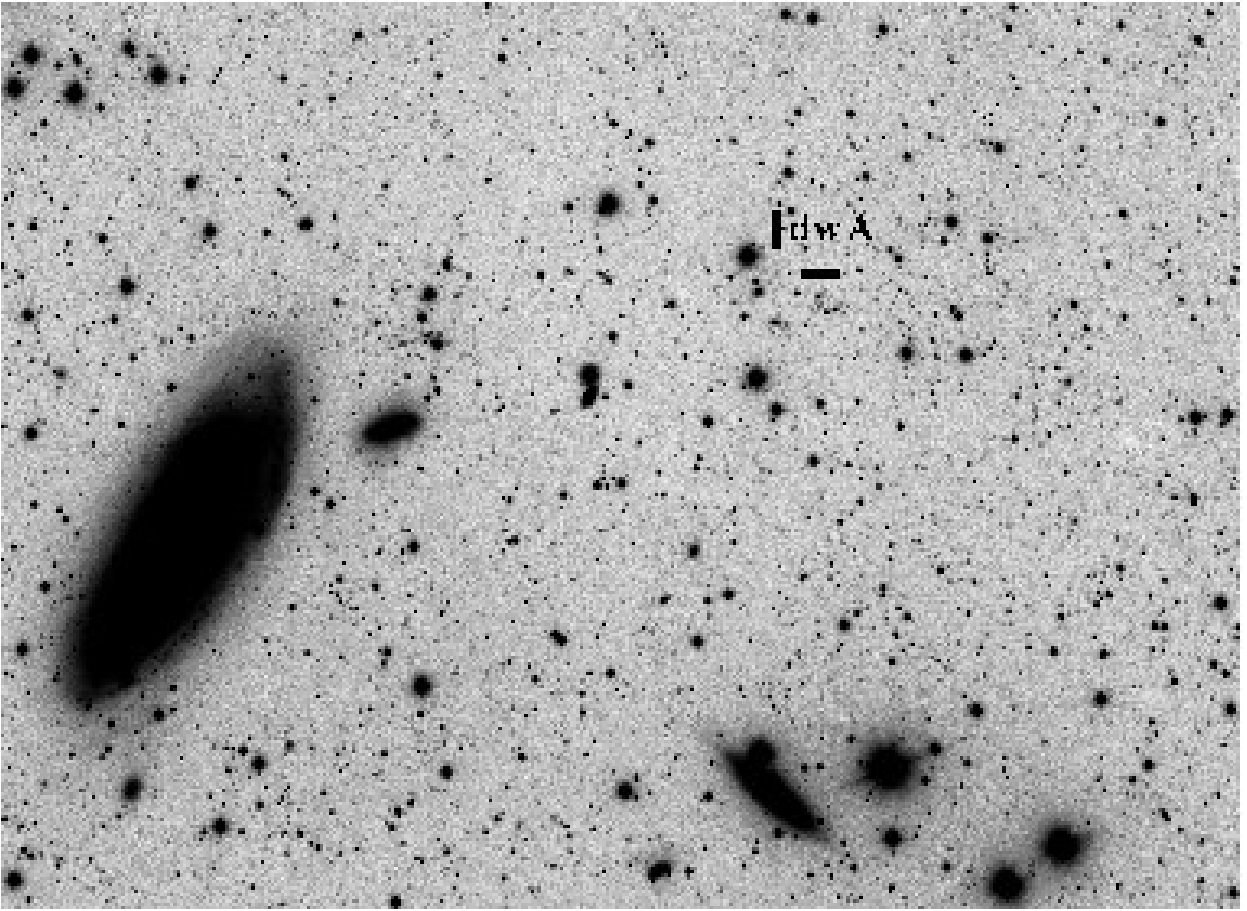}
\caption{The spiral galaxy NGC\,4258~$=$ M\,106 and its new
suspected satellite NGC\,4258\,dwA. The image was obtained by
 M.~Elvov with a 20-h exposure. The reproduced part of the image is $ 65\arcmin\times 48\arcmin$. North is
at the top, east is to the left.}
\end{figure*}

       \subsection{NGC\,4258 $=$ M\,106}
A careful search for faint satellites around the spiral galaxy
NGC\,4258 was complied by Kim
et~al.~\cite{kim2011:Karachentsev_n}.  For this purpose, the
authors used images taken with the   CFHT telescope MegaCam
detector, covering a field of  $1\fdg7\times 2\fdg0$. They found
sixteen M\,106 satellite candidates of  and presented for them the
surface photometry in the $g$ and $r$ bands. Two objects, S\,11
and S\,16, proved to be new compared to the previous
investigations of the M\,106
neighborhood~\cite{kar2007:Karachentsev_n}. Later, Spencer
et~al.~\cite{spe2014:Karachentsev_n} made
  radial-velocity measurements of the suspected satellites
of NGC\,4258.  This way they were trying to identify the physical
satellites of NGC\,4258 among the background galaxies. However,
the M\,106 group  is   located in a complicated region at the
equator of the Local Supercluster. Next to it there are two more
groups around the galaxies NGC\,4346 and NGC\,4157 with radial
velocities close to that of M\,106. The groups have average
distance estimates of 16.4~Mpc and 17.5~Mpc respectively, forming
the distant background for the   NGC\,4258 group ($D=7.83$~Mpc by
the Cepheid method).


The TBG group members obtained a few images of the neighborhood of
M\,106  with exposures ranging from $6^{\rm h}$ to $20^{\rm h}$.
In the area common with the field of view
of~\cite{kim2011:Karachentsev_n}, all the objects identified
in~\cite{kim2011:Karachentsev_n} as  suspected companions of
M\,106 were detected.
 Moreover, we have found a new LSB dwarf galaxy, indicated in
Fig.~6 as dwA. As Dr.~H.~Ann reported to us, in their image made
with MegaCam this object has not been noted, since it fell in the
gap between two CCD chips.
 In addition, outside the  MegaCam field  of view we found two more LSB objects  dwB and dwC, which look like very
probable companions of M\,106 (see {\tt
http://tbg.vdsastro.de}).\!\footnote{Near the southern edge of
Fig.~6, a distant background galaxy  NGC\,4217 is located. Its
radial velocity \mbox{$V_{\rm LG}=1084$~km\,s$^{-1}$},  and the
distance, according to NED, is 19.3~Mpc. On the northeastern edge
of NGC\,4217, a round LSB feature is visible, which may be a tidal
bulge in the disk of the spiral. If, however, this is a dwarf
companion of NGC\,4258, projected on the outskirts of NGC\,4217,
then its absolute magnitude $M_B=-11.5$, its linear diameter is
2.5~kpc, and the mean surface brightness ${\rm
SB}=26\fm8/\sq\arcsec$.}

        \subsection{NGC\,672/IC\,1727}

This tight pair of late-type spiral galaxies at a distance of
7.16~Mpc has three dwarf systems as its close satellites: KK\,13,
KK\,14, and KK\,15~\cite{kar2014a:Karachentsev_n}. In three images
of the neighborhood of the pair obtained by the  TBG members, we
have selected four probable satellite candidates. They are marked
on  the three panels of Fig.~7 by the letters A, B, and C. Another
candidate with the coordinates 014738.4+272620 is located in
contact with NGC\,672, and is the brightest \mbox{($B=18\fm7$)}
and the most compact. We have obtained its spectrum with the SAO
RAS 6-meter  telescope. The velocity of the object proved to be
very large,
\mbox{$V_h=29860\pm110$}~km\,s$^{-1}$~\cite{kar15:Karachentsev_n},
and we excluded it from the group. Three other  LSB galaxies look
more like close dwarfs. Radial-velocity measurements are required
to confirm their membership in the NGC\,672 group.

\begin{figure*}
 \setcaptionmargin{5mm} \onelinecaptionsfalse \captionstyle{normal}
 \vspace{2mm}
\includegraphics[width=0.41\textwidth]{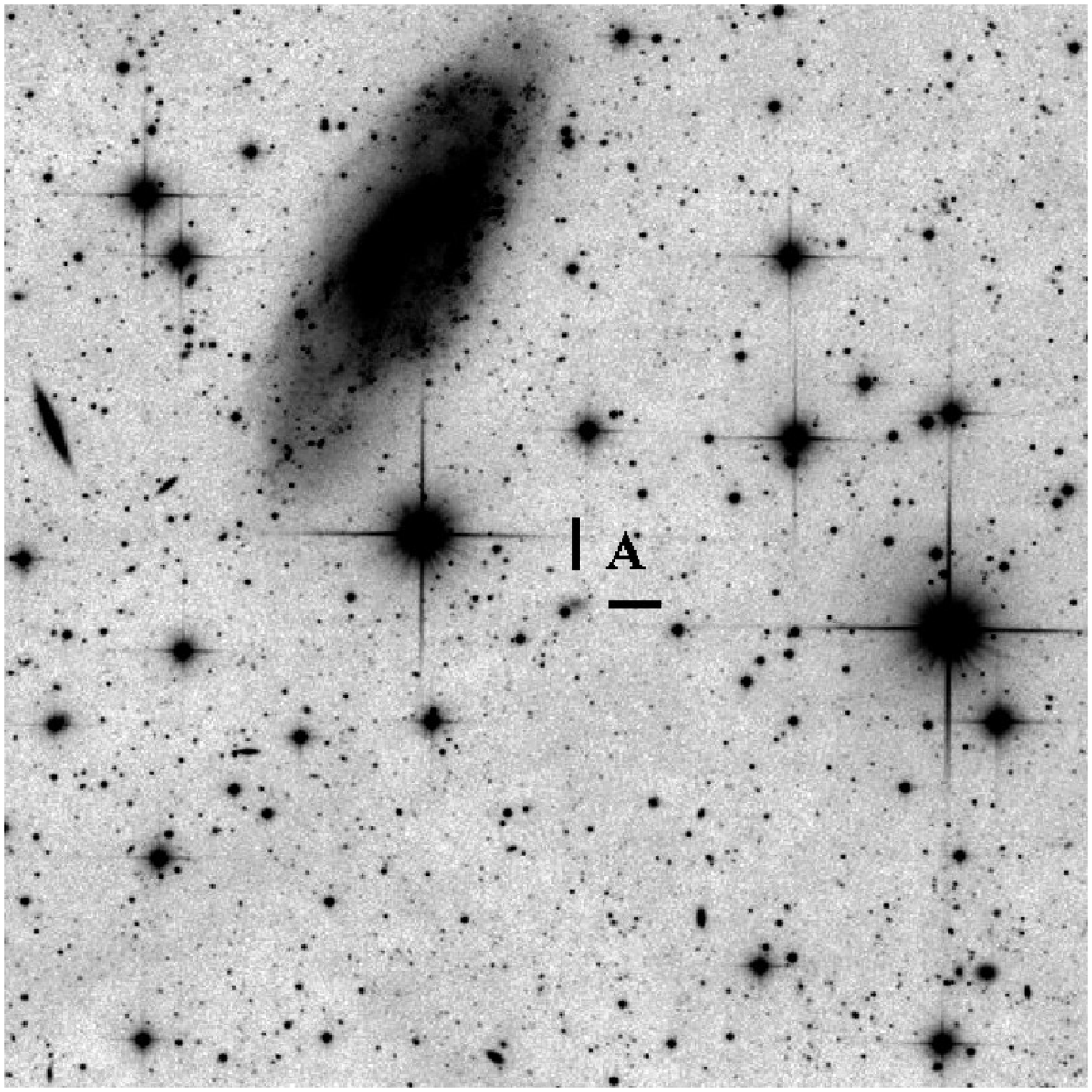}\hfill\includegraphics[width=0.57\textwidth]{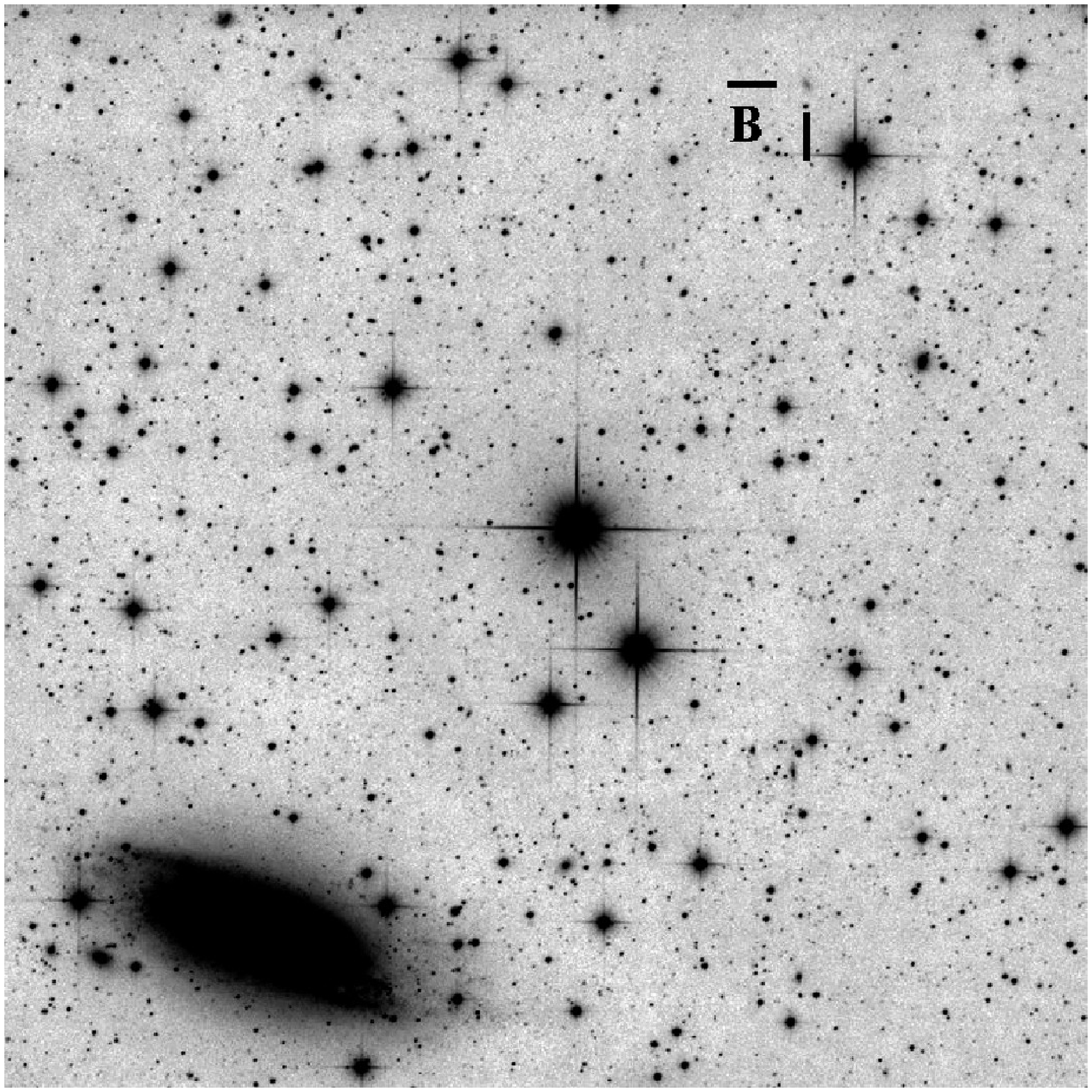}\\
 \hspace{-12mm}(a)\hspace{87mm}(b)\\
\vspace{2mm}
\includegraphics[width=0.33\textwidth]{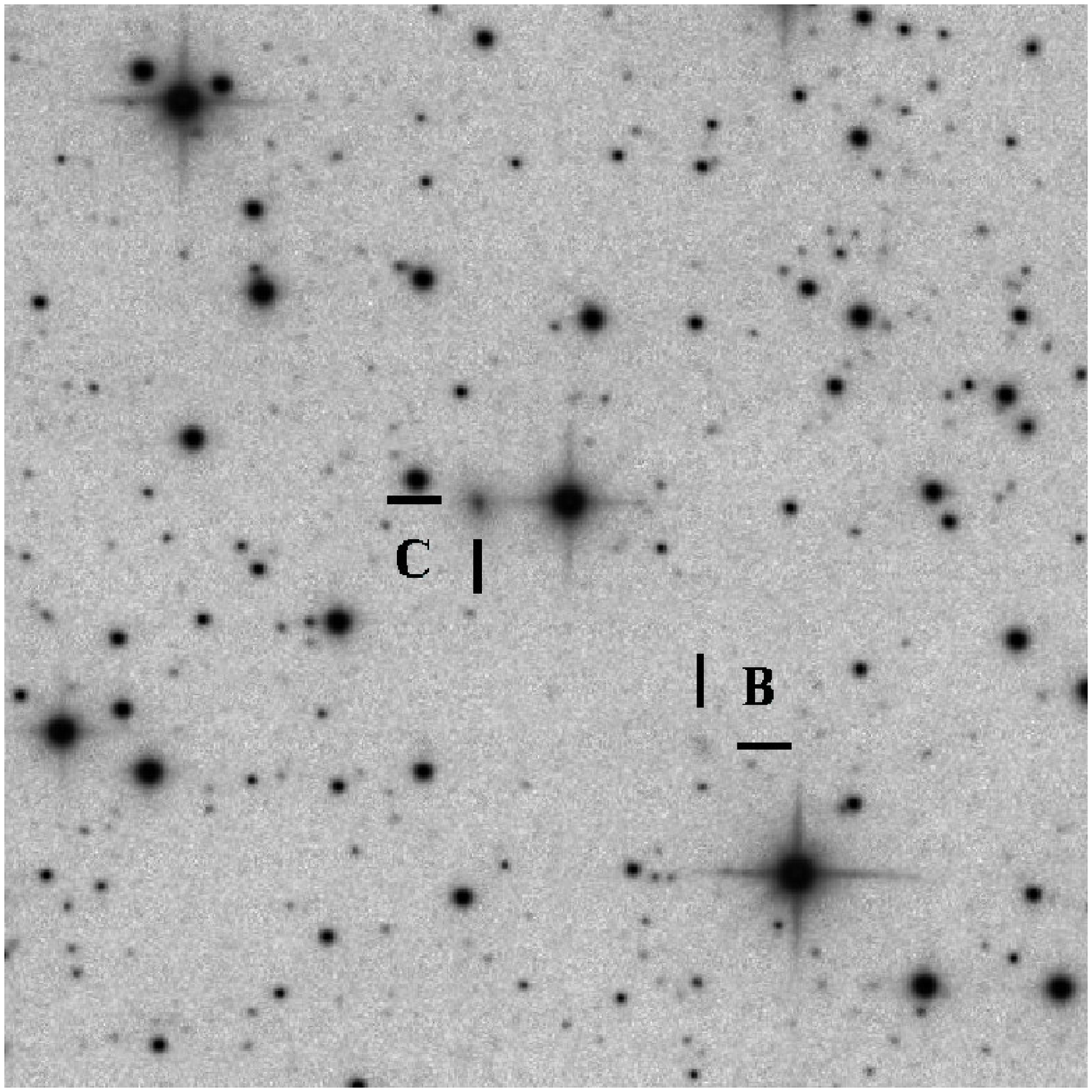}\\
(c)
 \caption{(a) The galaxy IC\,1727 and the dwarf galaxy
N\,672\,dwA (a $14\arcmin\times 14\arcmin$ part of an image
obtained by   R.~P\"{o}elz  with an exposure of $21^{\rm h}$).
(b)~The galaxy NGC\,672 and its assumed satellite dwB (a
$19\arcmin\times 19\arcmin$ fragment of the same image  by
R.~P\"{o}lzl). (c)~The area to the north of NGC\,672 with its two
supposed satellites NGC\,672\,dwB and NGC\,672\,dwC (a
$10\arcmin\times 10\arcmin$ fragment of an image obtained by
S.~K\"{u}ppers with an exposure of $5\fh3$). In all the images
north is at the top, east is to the left.}
\end{figure*}

        \subsection{NGC\,4618/NGC\,4625}

This is a pair of dwarf spirals at a distance of~7.9~Mpc with a
radial velocity difference of\linebreak   70~km\,s$^{-1}$.  The
spiral structure is distorted in both  galaxies, which indicates
their interaction. An \mbox{Im-type} dwarf galaxy  UGC\,7751 is
associated with this pair, and possibly another, fainter irregular
dwarf LV\,J1243+4127 too~\cite{huc2009:Karachentsev_n}.
Investigating an extended ultraviolet disk around NGC\,4625,
Gil~de~Paz\linebreak et~al.~\cite{gil2005:Karachentsev_n} noted
the presence $4\arcmin$ east of it of an LSB object, which they
named NGC\,4625\,A. The reproduction of an image of the
NGC\,4618/25 pair  and the dwarf galaxy NGC\,4625\,A obtained by
R.~Sparenberg is shown in Fig.~8. We have included NGC\,4625\,A in
the list of objects for measuring radial velocities, assuming that
it can prove to be a physical companion of the NGC\,4618/25 pair.

\begin{figure}
 \setcaptionmargin{5mm} \onelinecaptionsfalse \captionstyle{normal}
 \vspace{2mm}
\includegraphics[width=\columnwidth]{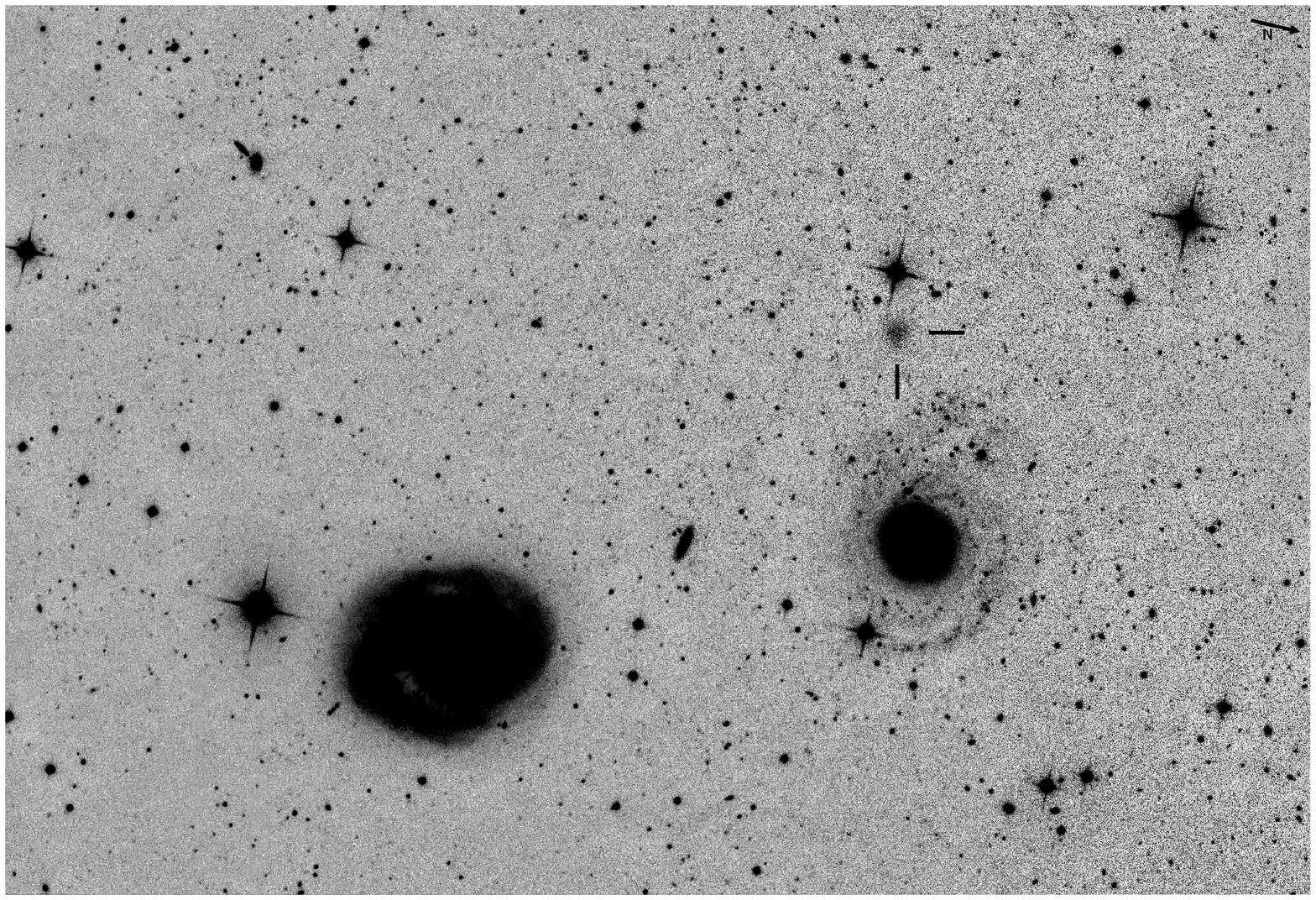}
\caption{An interacting pair of galaxies NGC\,4618 (left) and
NGC\,4625, and a dwarf galaxy NGC\,4626\,A, marked by the arrows.
A $23\arcmin\times 16\arcmin$ fragment of an  image  obtained by
R.~Sparenberg  on a 44-inch telescope with an exposure of $3\fh5$.
The arrow in the upper right corner indicates the north
direction.}
\end{figure}

        \subsection{NGC\,1156 $=$ KIG\,121}

 This isolated  Magellanic type dwarf galaxy  with an apparent
magnitude   $B=12\fm3$  and radial velocity $V_h=373$~km\,s$^{-1}$
is located in the zone of significant ($A_b=0\fm97$) Galactic
extinction.  The image of its surroundings with a $46\arcmin\times
34\arcmin$ field of view
 was obtained by P.~Hochleitner on a
36~cm diameter telescope with an exposure of $11^{\rm h}$.
Reflection nebulae (cirri), the presence of which imposes a limit
on the detection of LSB galaxies, are seen almost over the entire
area of the image. Nevertheless, two LSB objects, dw1 and dw2,
which can be classified as satellites of NGC\,1156, are present in
the image (Fig.~9). The first one of them is located in the halo
of a bright star SAO\,75679. Both objects are barely discernible
in the photographic sky survey DSS2.

Note that the vicinity of NGC\,1156 was investigated by Minchin
et~al.~\cite{min2010:Karachentsev_n} within the deep H\,I AGES
survey using the  Arecibo radio telescope. These observations led
to the discovery of an irregular dwarf galaxy AGES\,J030039+254656
with a radial velocity
 $V_h=308$~km\,s$^{-1}$ and  apparent magnitude \mbox{$B=18\fm1$},
 which is located  to the north, beyond the scope of our image.
Two NGC\,1156 satellite candidates that we have discovered look
about two magnitudes fainter than the AGES satellite. Obviously,
their H\,I fluxes could be below the AGES survey limit.

\subsection{NGC\,2903, NGC\,3239, NGC\,4214, NGC\,5585}

Apart from the nine galaxies   mentioned above, we have obtained
long-exposure images for  four other galaxies in the Local Volume.
 However, the search for new candidate satellites of these galaxies proved to be unsuccessful.

 The massive spiral galaxy NGC\,2903 is listed in the catalog of isolated galaxies~\cite{kara1973:Karachentsev_n}.
Although it has four tiny satellites, due to their small size this
is not inconsistent with the catalog isolation criterion. Just
like in the case of NGC\,2683, this galaxy is on the front edge of
the nearby Gemini--Leo Void, whose center lies about 18.4~Mpc from
us, and whose radius is 7.5~Mpc. Deep images of the vicinity of
NGC\,2903 made by M.~Blauensteiner, M.~Elvov, and S.K\"{u}ppers
did not reveal any new companion candidates of the massive spiral.


An interactive pair of merging galaxies \linebreak
\mbox{NGC\,3239~$=$ Arp\,263~$=$ VV\,095}  contains a scattering
of star formation centers. Two curved tails extending to the south
of the main body of the galaxy make it similar to the Greek letter
``$\pi$.\!'' Around this violently interacting system one would
expect to find a multitude of small irregular satellites formed by
the fragmentation of tidal tails. However, they have not been
found in the image made by B.~Hubl with an exposure of $32^{\rm
h}$.

A magellanic-type dwarf galaxy NGC\,4214 at a distance of
2.94~Mpc~\cite{kar2013:Karachentsev_n} is in the stage of violent
star formation. Next to it there is a dwarf spheroidal galaxy
KDG\,90, the distance to which,
2.86~Mpc~\cite{kar2013:Karachentsev_n}, indicates the physical
connection with NGC\,4214. However, the images obtained by
G.~Willems and G.~Kerschhuber with an exposure of $19\fh1$  do not
show any obvious signs of mutual perturbations in these galaxies,
apart from  a faint diffuse protrusion  on the southwestern
periphery of NGC\,4214. No new NGC\,4214 satellite candidates have
either been found.

A dwarf Sm spiral NGC\,5585 at a distance of 5.7~Mpc is a
distant  satellite of a giant spiral M\,101. The periphery of the
galaxy looks quite regular, without disturbances. The image
obtained by B.~Hubl with an exposure of $15^{\rm h}$ did not
   reveal any satellite candidates of this galaxy in the $30\arcmin\times 30\arcmin$ field.
The reproductions of the said images and the comments to them can
be found on \mbox{\tt http://tbg.vdsastro.de}.

         \section{DISCUSSION}

The  main features of the thirteen nearby spiral galaxies that we
have observed with long exposures are presented in Table~1. Its
columns contain: (1)~the names of the galaxies, ranked by right
ascension; (2)~the coordinates of the galaxy for the epoch
J2000.0;
   (3)~the morphological type according to de~Vaucouleurs' scale;
   (4)~the radial velocity in~km\,s$^{-1}$ relative to the centroid of the Local Group;
(5)~the distance to the galaxies in Mpc and the applied method of
distance determination: ``cep''---the Cepheid luminosity method,
``rgb''---by the luminosity of red giant branch stars,
``sbf''---by surface brightness fluctuations, ``tf''---by the
   Tully--Fisher relation between the amplitude of  rotation of a galaxy and its luminosity, ``bs''---by the luminosity of the brightest stars;
(6)~the absolute magnitude of the galaxy in the  $B$~band adjusted
for the Galactic~\cite{sch1998:Karachentsev_n} and internal
extinction; (7)~the linear Holmberg diameter of the galaxy in~kpc;
(8,~9)~the logarithm of the
   stellar mass  and the mass of neutral hydrogen in  solar mass units;
(10)~the logarithm of the stellar mass density within the 1~Mpc
radius sphere around the galaxy, taken in relation to the average
cosmic stellar mass density; (11)~the number of  known satellites
of the galaxy within the ``zero-velocity radius'' sphere around it
plus the number of new  satellite candidates that we found and
discussed above; (12)~the signs of disturbances on the periphery
of the galaxy if any of them can be seen in our deep images.

\begin{figure}
 \setcaptionmargin{5mm} \onelinecaptionsfalse \captionstyle{normal}
 \vspace{2mm}
\includegraphics[width=0.855\columnwidth]{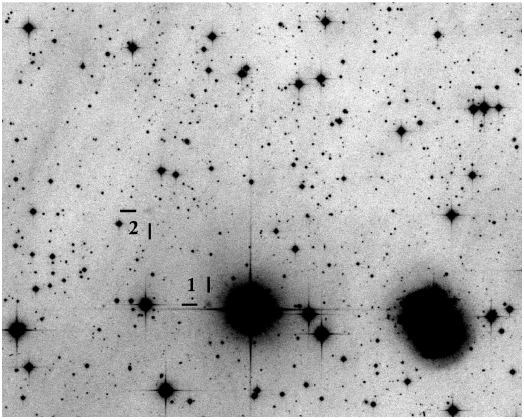}
\caption{An isolated galaxy NGC\,1156 in the lower right corner
and two of its suspected satellites dw1 and dw2. A
$19\arcmin\times 15\arcmin$ fragment of an image obtained by
P.~Hochleitner with an exposure of $11^{\rm h}$. North is at the
top, east is to the left.}
\end{figure}

\renewcommand{\baselinestretch}{1.2}
\setcaptionmargin{0mm}
 \onelinecaptionstrue
 \captionstyle{normal}
\begin{table*}[b]
\caption{Nearby massive galaxies observed by the TBG group}
\medskip
\begin{tabular}{l|c|c|c|l|c|r|r|c|r|r@{$\,+\,$}l|l}
 \hline
 \multicolumn{1}{c|}{Name}  & RA (2000.0) Dec & $T$ & $V_{\rm LG}$,    & \multicolumn{1}{c|}{$D$,~Mpc;}  &   $M_B$, & \multicolumn{1}{c|}{$A_H$,} & \multicolumn{1}{c|}{$\log M^*$,}   &  $\log M_{\rm HI}$, & \multicolumn{1}{c|}{$\Theta_j$} & \multicolumn{2}{c|}{$N_{\rm sat}$} & \multicolumn{1}{c}{Periphery}\\
                                             &                                  &                      & km\,s$^{-1}$ & \multicolumn{1}{c|}{method}     &    mag   & \multicolumn{1}{c|}{kpc}    & \multicolumn{1}{c|}{$[M_{\odot}]$} &  $[M_{\odot}]$        &                                                  & \multicolumn{2}{c|}{}                               & \multicolumn{1}{c}{shape}\\
\hline
 \multicolumn{1}{c|}{(1)}                    &       (2)                        &(3)                   & (4)          & \multicolumn{1}{c|}{(5)}        &  (6)     & \multicolumn{1}{c|}{(7)}    & \multicolumn{1}{c|}{(8)}           &   (9)                 & \multicolumn{1}{c|}{(10)}                        & \multicolumn{2}{c|}{(11)}                           & \multicolumn{1}{c}{(12)}\\
\hline
 NGC\,672  & 014753.2+272601 & 6 & 626 & 7.16 rgb & $-18.76$ & 15.7 & 10.22~~ &  9.23  & $0.2$   & 3  & 3  & regular\\
 NGC\,891  & 022232.8+422048 & 3 & 736 & 9.97 sbf & $-20.58$ & 38.7 & 10.98~~ &  9.66  & $-0.1$  & 4  & 2  & regular\\
 NGC\,1156 & 025942.4+251415 & 8 & 507 & 7.80 bs  & $-18.14$ &  9.2 &  9.31~~ &  8.82  & $-1.5$  & 1  & 2  & NE extention\\
 NGC\,2683 & 085240.9+332502 & 3 & 365 & 9.36 rgb & $-20.78$ & 35.7 & 10.76~~ &  9.10  & $-1.4$  & 2  & 2  & regular\\
 NGC\,2903 & 093209.6+213002 & 4 & 443 & 8.87 bs  & $-20.89$ & 32.4 & 10.82~~ &  9.44  & $-0.8$  & 4  & 0  & regular\\
 NGC\,3239 & 102504.9+170949 & 8 & 623 & 7.90 tf  & $-18.09$ & 11.6 &  9.52~~ &  8.89  & $-0.9$  & 0  & 0  & faint halo\\
 NGC\,3344 & 104330.2+245525 & 4 & 500 & 9.83 rgb & $-19.72$ & 22.2 & 10.33~~ &  9.44  & $-0.8$  & 0  & 1  & regular? \\
 NGC\,4214 & 121538.9+361939 & 8 & 295 & 2.94 rgb & $-17.20$ &  7.3 &  9.00~~ &  8.48  & $-0.7$  & 3  & 0  & SSW protrusion?\\
 NGC\,4258 & 121857.5+471814 & 4 & 506 & 7.83 cep & $-21.20$ & 41.5 & 10.94~~ &  9.64  & $1.0$   & 19 & 3  & twisted\\
 NGC\,4618 & 124132.8+410903 & 6 & 576 & 7.90 tf  & $-18.33$ &  9.7 &  9.65~~ &  8.90  & $0.5$   & 2  & 1  & N, E protrusions\\
 NGC\,4631 & 124208.0+323229 & 7 & 581 & 7.38 rgb & $-20.28$ & 33.7 & 10.49~~ &  9.72  & $1.0$   & 5  & 3  & tails\\
 M\,101    & 140312.8+542102 & 6 & 378 & 7.38 cep & $-21.12$ & 65.2 & 10.85~~ &  9.91  & $0.2$   & 6  & 10 & asymmetric \\
 NGC\,5585 & 141948.3+564349 & 7 & 457 & 5.70 bs  & $-17.81$ &  9.7 &  9.03~~ &  8.82  & $-3.0$  & 0  & 0  & regular\\
\hline
\end{tabular}
\end{table*}
\renewcommand{\baselinestretch}{1.0}

Most of the   Table~1 data are taken from the
UNGC~\cite{kar2013:Karachentsev_n}, which gives evaluations of the
parameters used.
 Some distance estimates  have been updated with recent observations of the Hubble Space Telescope.


Despite the small statistics, the  Table~1 data show the presence
of a positive correlation between the luminosity of a galaxy (its
linear diameter, mass) and the number of satellites around it. The
average density of the environment also affects the amount of
physical satellites. For example,  two galaxies in the region of a
cosmic void, NGC\,2683 and NGC\,2903, have six satellites and two
new candidates, while the two galaxies in the dense environment,
NGC\,4258 and NGC\,4631 with approximately the same luminosity,
have twenty-four  physical satellites and six new candidates. The
noted trends suggest near which galaxies there is a chance to find
the greatest number of new satellites during a further search.

The distant periphery of Andromeda-type massive galaxies preserves
in itself in the form of  faint stellar streams relic signs of
interaction with neighbors  or traces of the capture of satellites
which happened a few
  billion years ago.
The last column of Table~1 provides indications on the possible
events of this kind
   in the history of thirteen nearby high luminosity galaxies.

\renewcommand{\baselinestretch}{1.15}

\setcaptionmargin{0mm}
 \onelinecaptionstrue
 \captionstyle{normal}
\begin{table*}[b]
\caption{New LSB satellite candidates around nearby massive
galaxies}
\medskip
\begin{tabular}{l|c|c|c|c|c|r|r|r|c|l}
\hline
\multicolumn{1}{c|}{Name} & RA (2000.0) Dec & $T$  & $B$, &   $a$, & ${\rm SB}$ & \multicolumn{1}{c|}{$r_p$,} & $R_p$,& \multicolumn{1}{c|}{$M_B$,} & $A$,  & \multicolumn{1}{c}{Note}\\
                                           &                                  &                       &  mag & arcmin &                       & \multicolumn{1}{c|}{arcmin} & kpc  & \multicolumn{1}{c|}{mag}    & kpc   &\\
\hline
 \multicolumn{1}{c|}{(1)}                  & (2)                              &  (3)                  &  (4) &  (5)   & (6)                   & \multicolumn{1}{c|}{(7)}    & (8)  & \multicolumn{1}{c|}{(9)}    & (10)  & \multicolumn{1}{c}{(11)}\\
\hline
NGC\,672\,dwB   & 014711.1+274100 & Ir-VL  &  21.0 &  0.20 & 25.8 & 17.7~~ & 37  & $-8.6$  & 0.42 & \\
NGC\,672\,dwA   & 014719.1+271516 & Ir-L   &  19.8 &  0.26 & 25.2 & 13.1~~ & 27  & $-9.8$  & 0.54 & GALEX\\
NGC\,672\,dwC   & 014720.4+274324 & Sph-L  &  18.7 &  0.40 & 25.0 & 18.9~~ & 39  & $-10.9$ & 0.83 & \\
NGC\,891\,dwA   & 022112.4+422150 & Tr-L   &  17.9 &  0.76 & 25.7 & 14.9~~ & 43  & $-12.3$ & 2.20 & [TT09]25\\
NGC\,891\,dwB   & 022254.7+424245 & Ir-VL  &  18.9 &  1.16 & 27.6 & 22.4~~ & 65  & $-11.3$ & 3.36 & [TT09]30\\
NGC\,1156\,dw1  & 030018.2+251456 & Ir-L   &  19.6 &  0.38 & 25.2 & 8.1~~  & 18  & $-10.8$ & 0.86 & \\
NGC\,1156\,dw2  & 030028.0+251817 & Ir-VL  &  20.0 &  0.38 & 25.6 & 11.1~~ & 25  & $-10.4$ & 0.86 & GALEX\\
NGC\,2683\,dw1  & 085326.8+331820 & Ir-L   &  19.0 &  0.40 & 25.5 & 11.7~~ & 32  & $-11.0$ & 1.09 & GALEX\\
NGC\,2683\,dw2  & 085420.5+331458 & Sph-VL &  19.6 &  0.40 & 26.1 & 23.1~~ & 63  & $-10.4$ & 1.09 & \\
NGC\,3344\,dw1  & 104244.0+250130 & Ir-VL  &  20.0 &  0.30 & 26.0 & 11.9~~ & 34  & $-10.1$ & 0.86 & \\
NGC\,4258\,dwC  & 121026.8+464449 & Sph-L  &  19.0 &  0.27 & 24.7 & 93.3~~ & 212 & $-10.5$ & 0.61 & \\
NGC\,4258\,dwA  & 121551.0+473256 & Ir-L   &  19.0 &  0.43 & 25.7 & 34.8~~ & 79  & $-10.5$ & 0.98 & \\
NGC\,4258\,dwB  & 122410.9+470723 & Sph-L  &  18.3 &  0.45 & 25.1 & 54.6~~ & 124 & $-11.2$ & 1.02 & BTS134\\
NGC\,4631\,dw1  & 124057.0+324733 & Ir-VL  &  16.1 &  2.20 & 26.4 & 21.3~~ & 46  & $-13.3$ & 4.72 & GALEX\\
NGC\,4631\,dw2  & 124206.8+323715 & Ir-VL  &  18.5 &  0.90 & 26.8 & 4.8~~  & 10  & $-10.9$ & 1.93 & GALEX\\
NGC\,4625\,A    & 124211.0+411510 & Tr-L   &  18.6 &  0.45 & 25.4 & 9.4~~  & 22  & $-11.0$ & 1.03 & \\
NGC\,4631\,dw3  & 124252.5+322735 & Sph-VL &  19.7 &  0.60 & 27.1 & 10.6~~ & 23  & $-9.7$  & 1.29 & \\
M\,101\,DF3     & 140305.7+533656 & Sph-VL &  17.9 &  1.00 & 26.5 & 44.1~~ & 95  & $-11.5$ & 2.15 & \\
M\,101\,DF1     & 140345.0+535640 & Ir-L   &  18.9 &  0.47 & 25.8 & 23.9~~ & 51  & $-10.5$ & 1.01 & \\
M\,101\,dwD     & 140424.6+531619 & Sph-VL &  19.2 &  0.38 & 25.7 & 65.6~~ & 141 & $-10.2$ & 0.81 & \\
M\,101\,dwC     & 140518.0+545356 & Tr-VL  &  20.2 &  0.30 & 26.2 & 37.6~~ & 81  & $-9.2$  & 0.64 & \\
M\,101\,DF7     & 140548.3+550758 & Sph-XL &  20.4 &  0.67 & 28.1 & 52.0~~ & 117 & $-9.0$  & 1.44 & \\
M\,101\,dwA     & 140650.2+534432 & Sph-L  &  19.2 &  0.36 & 25.6 & 45.3~~ & 97  & $-10.2$ & 0.77 & \\
M\,101\,DF4     & 140733.4+544236 & Ir-XL  &  18.8 &  0.93 & 27.2 & 43.5~~ & 93  & $-10.6$ & 1.99 & \\
M\,101\,DF6     & 140819.0+551124 & Ir-XL  &  20.1 &  0.73 & 28.0 & 67.2~~ & 144 & $-9.3$  & 1.57 & \\
M\,101\,DF2     & 140837.5+541931 & Sph-L  &  19.8 &  0.33 & 26.0 & 47.1~~ & 101 & $-9.6$  & 0.71 & \\
M\,101\,dwB     & 140843.1+550957 & Sph-VL &  20.1 &  0.30 & 26.1 & 68.0~~ & 146 & $-9.3$  & 0.64 & \\
\hline
 Mean           &                 &        &  19.2 &  0.57 & 26.1 & 32.4~~ & 73  & -10.4   & 1.31 & \\
\hline
\end{tabular}
\end{table*}
\renewcommand{\baselinestretch}{1.0}

A summary of the data on twenty-seven detected satellite
candidates of the thirteen  nearby spiral galaxies is given in
Table~2. Its columns contain: (1)~the name of the dwarf galaxy;
(2)~the galaxy coordinates for the epoch 2000.0; (3)~the
morphological type of the dwarf: irregular (Ir), spheroidal (Sph),
or transition (Tr), with a visual estimation of surface
brightness: low (L),  very low (VL), or extremely low (XL);
(4)~the apparent $B$~magnitude estimated by eye via comparing with
other dwarf objects of similar structure and known photometry; in
some of the brightest  objects the $B$~magnitude is    determined
from the SDSS survey $g$ and $r$ magnitudes, the measure of
inaccuracy of our estimates is about
  $0\fm5$; (5)~the maximum apparent angular diameter $a$ in arcminutes;
(6)~the surface brightness~\mbox{${\rm SB}=B^c+5\log a\arcmin +
8.63$}
      in magnitudes per square arcsecond, where the
  $B$~magnitude is corrected for the Galactic extinction;
 (7,~8)~the projected distance of the satellite from the main galaxy in
arcminutes and in kpc; (9, 10)~the absolute magnitude and linear
diameter of the satellite; (11)~whether the object has a FUV flux
from the GALEX  survey or an alias. The last line of the table
shows the average values of the above parameters.


We  can  see from these data that the absolute magnitudes of the
suspected new satellites around nearby spirals are confined in the
range from
  \mbox{$-8\fm6$} to $-13\fm3$ with an average value of $-10\fm4$.
Linear diameters of the satellites lie in the range from 0.4~kpc
to 4.7~kpc with an average of 1.3~kpc. The range of both
parameters is typical for the dwarf satellites of M\,31 and M\,81.
The average surface brightness of the discovered dwarfs
\mbox{$\langle {\rm SB}\rangle = 26\fm1/\sq\arcsec$}  barely
exceeds the Holmberg isophote brightness of $26\fm5/\sq\arcsec$,
which is the detection limit  for low-contrast objects observed
with photographic emulsions.


The average projected separation of  new satellites from their
host galaxies is 73~kpc. This value is \mbox{3--4 times}  smaller
than the characteristic radius  of an entourage of  dwarfs around
a Milky Way--type massive galaxy. It is obvious that such a
difference is due to
  the small field of view of the used telescopes, which is comparable to the average angular separation of the satellites
\mbox{$\langle r_p\rangle = 32\farcm4$}. It hence follows that  a
significant number of low and very low surface brightness dwarf
satellites can still be found around massive Local Volume galaxies
in the field of view with a radius of \mbox{$1\degr$--$2\degr$}.
However, their search would require a lot more observing time of
the telescopes in use or attraction  of new astronomical
\mbox{CCD} imaging enthusiasts to this program.

         \section{CLOSING REMARKS}

Detection of ultra-faint dwarf satellites around  nearby massive
galaxies is of  important cosmological significance.
  Successful search for such objects in the Local Group around the Milky Way and the Andromeda Nebula (M\,31)
attracted great interest to the features of their spatial  distribution and
kinematics~\cite{kro2005:Karachentsev_n,iba2013b:Karachentsev_n,sha2013:Karachentsev_n}.

Many believe that ultra-faint dwarf galaxies are among the
``darkest'' objects in the Universe and in this sense they can
serve as a natural laboratory to study the nature of dark matter.


The program of detection of faint dwarf galaxies with small
telescopes should obviously  be accompanied by  systematic radial
velocity measurements of new    LSB  objects to confirm their
physical connection with  massive galaxies. Radial velocity
measurements as well as the  studies of the photometric structure
of  new dwarf galaxies require the capabilities of large
telescopes. It remains to add that extending the program to the
southern sky objects is a  quite evident and topical task.

        \begin{acknowledgements}
The authors are grateful to S.~S.~Kaisin,\linebreak D.~I.~Makarov,
M.~E.~Sharina, Yu~A.~Perepelitsyna, and E.~S.~Safonova, who took
part in the observations at the 6-m telescope and data processing.
This work was supported by Russian Science Foundation grant
No.~14-12-00695. Observations on the \mbox{6-m} telescope of the
Special Astrophysical Observatory are held with financial support
from the Ministry of Education and Science of the Russian
Federation (contract No.~14.619.21.0004, project identifier
RFMEFI61914X0004).
        \end{acknowledgements}

        {}
\end{document}